\documentclass[a4paper]{jpconf}
\usepackage{graphicx}
\usepackage{multirow}

\usepackage{aliascnt}

\usepackage{pgf}
\usepackage{csquotes}
\usepackage{pdfpages}
\usepackage{amsmath}
\usepackage[amsmath,thmmarks]{ntheorem}

\usepackage{cleveref}

\usepackage[normalem]{ulem}
\usepackage{color}

\begin{document}
\title{In-medium $J/\psi$ { mass shift by the} $D$ meson loop effect}

\author{Rahul Chhabra$^1$ and Kazuo {Tsushima}$^2$}

\address{$^1$Department of Physics, Dr. B R Ambedkar National Institute of Technology Jalandhar,
Jalandhar -- 144011, Punjab, India}

\address{$^2$Laboratorio de Fısica Teorica e Computacional-LFTC
Universidade Cruzeiro do Sul and Universidade Cidade de Sao Paulo (UNICID)
01506-000, Sao Paulo, SP, Brazil}
\ead{rahulchhabra@ymail.com}

\begin{abstract}
By an effective Lagrangian plus QCD sum-rule approach, 
we investigate the mass shift of the $J/\psi$ state in medium, 
in symmetric nuclear matter with zero and finite temperature, 
and cold strange matter. 
The in-medium mass of the $J/\psi$ state is evaluated through 
the intermediate pseudoscalar $D$-meson loop for the $J/\psi$ self-energy.  
The effect of medium is incorporated through the in-medium mass of $D$ meson 
calculated using chiral SU(3) model plus QCD sum-rule approach. 
The self energy loop { integral is} regularized using the phenomenological 
form factor of the dipole form.
We compare our results with some of { the results} in the literature.   
The present results should be helpful to understand {better} the expected data from  
heavy ion collision experiments, such as CBM and PANDA.
\end{abstract}

\section{Introduction}
The study of in-medium hadron properties is very important to understand better the 
data from heavy ion collision (HIC) experiments. 
For example, at facility of antiproton and ion research (FAIR), 
charmed hadrons will be produced copiously by the annihilation of antiproton on nuclei. 
The { studies} of charmed hadron as well as charmonium properties in medium 
will advance { in understanding} the interaction of these mesons with the nucleons. 
In particular, their properties in a strange hadronic medium will 
add interesting, extra and new information. 
For example, $J/\psi$ $(c\bar{c})$ suppression is known as one of the signatures of 
the quark gluon plasma (QGP) production in HICs. 
It was suggested first by Matsui and Satz that the drop in the yield of the $J/\psi$ 
state in HICs is due to the color screening effect, 
and this should be considered as a signature of the production of QGP~\cite{mats}. 
Important results in favor of $J/\psi$ suppression were indeed observed at CERN SPS 
and RHIC experiments~\cite{na50,na60,phenix}. 
However, in Refs.~\cite{domi2010, mocsy2007,asakawa2004} 
it was observed that the heavy quark bound states can survive in the deconfined QGP phase.   
Therefore, further studies on the modification of the $J/\psi$ state in medium 
will add extra important information on the production of QGP in heavy ion experiments. 
Moreover, the charmonium interaction with the nucleon and hadronic 
medium was investigated in Refs.~\cite{kling,hayaj,sug}, 
and they predict the negative shift in the mass (around 7-8 MeV) of $J/\psi$ meson 
by using QCD sum-rule approach. 
Also, the charmonium nucleon interaction and  dissociation cross section 
of $J/\psi$ state were studied using one-boson exchange model~\cite{sibi}, 
quark model~\cite{hil}, and effective Lagrangian approach~\cite{liu,liu2}. 
In the present investigation, we study the impact of the pseudoscalar $D$-meson 
loop contribution on the mass shift of $J/\psi$ state in hot and strange hadronic medium. 
The medium effects are incorporated through the in-medium  
{ modified mass} of $D$ meson, calculated using chiral SU(3) model plus QCD sum-rule 
approach~\cite{rahul}.

\section{Methodology}
We use the phenomenological effective Lagrangian density at hadronic level for 
the vertices $J/\psi$-$DD$~\cite{ko,arv}, 

\begin{align}
{\cal{L}}_{int}  = ig_{J/\psi DD} {(J/ \psi)}^{\mu} [\bar{D}(\partial _{\mu} D) - (\partial _{\mu} 
\bar{D})D ] + 
g^2_{J/\psi DD} {J/ \psi}_{\mu} {(J/ \psi)}^{\mu} \bar{D}D ,
\end{align}
with conventions,  $D$ =
$\begin{pmatrix}
{ D^+} \\
{ D^0}
\end{pmatrix}$,
and $\bar{D}$ = $\begin{pmatrix}
\bar{D}^0 & D^-
\end{pmatrix}$.

Here the Lagrangian density is an SU(4) extension of the light-flavor chiral-symmetric Lagrangians  
of the pseudoscalar mesons. However, SU(4) flavor symmetry is strongly broken in nature. 
Therefore, we use experimental masses of the open charmed mesons and $J/\psi$ meson. 
In addition, we employ the empirically extracted coupling constants of mesons. 
The shift in the mass of $J/\psi$ meson is defined as 
$\Delta m_{J/\psi}$ = $m^*_{J/\psi}$ - $m_{J/\psi}$, 
where, $m^*_{J/\psi}$ represents the in-medium mass of $J/\psi$ meson, 
while $m_{J/\psi}$ is the free-space mass. 

Furthermore, using the non-gauged Lagrangian density for the $J/\psi$-$DD$ vertices, 
we get the expression for the $J/\psi$ self-energy by the $D$-meson loop contribution~\cite{tsu}:
\begin{align}
\Sigma_{DD}(m^2_{J/\psi})=\frac{-g^2_{J/\psi DD}}{3\pi^2} \int dq \frac{q^2}{(q^2+m^2_D)^{1/2}}
\Big(\frac{q^2}{(q^2+m^2_D)^2-m^2_\psi/4}\Big)F_{DD}(q^2).
\end{align} 
In the above { $q \equiv |\vec{q}|$,} and $F_{DD}(q^2)$ represents 
the {square of} vertex form factor { ($u(q^2)$) multiplied to the two vertices,} 
and we use a dipole form~\cite{tsu} {with $\omega_D(q) = (q^2 + m_D^2)^{1/2}$,}
\begin{align}
F_{DD}(q^2) = u^2(q^2) = \left(\frac{\Lambda^2_D + m^2_{J/\psi}}{\Lambda^2_D + 4\omega^2_D(q)} \right)^4 ,
\end{align}
where, $\Lambda_D$ is the cut off mass, and  $\Lambda_D = 1000$ MeV is used in this study.  
The bare mass $(m^0_{J/\psi})$ is calculated from the free-space values of the 
$J/\psi$ and $D$ mesons as,
\begin{align}\label{mpsi}
{m^2_{J/\psi}} = (m^0_{J/\psi})^2+\Sigma(k^2=m^2_{J/\psi}),
\end{align}
where, $m^0_{J/\psi}$ is the bare mass and $\Sigma(k^2=m^2_{J/\psi})$ is 
the total $J/\psi$ self-energy obtained from the contribution from the $DD$ loop. 
To study the in-medium mass of the $J/\psi$, 
we use the in-medium mass of $D$ meson calculated using chiral SU(3) model plus QCD 
sum-rule approach~\cite{rahul} as input in the in-medium $J/\psi$ self-energy, 
where the bare mass is fixed in free space.

\begin{figure}
\begin{center}
\includegraphics[width=10cm]{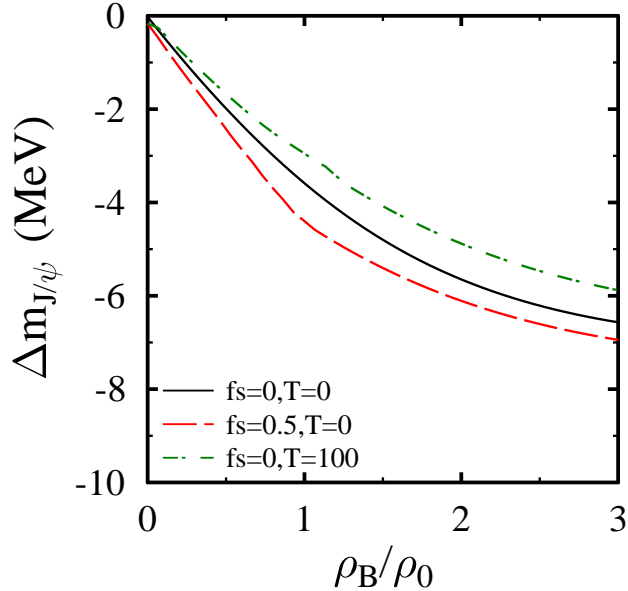}
\caption{\label{fig}
In-medium $J/\psi$ meson mass calculated by the $D$-meson loop effect, 
as a function of baryonic density $\rho_B/\rho_0$ ($\rho_0 = 0.15$ fm$^{-3}$) at finite temperature 
($T = 100$ MeV) and strangeness fraction, $f_s = 0.5$.
}
\end{center}
\end{figure}

\section{Results and discussion}

In the present study, we use the value of the coupling constant  
$g_{J/{\psi} DD} = 7.64$~\cite{tsu}. 
The free-space values of $m_{J/\psi}$ and $m_D$ 
used are 3096 and 1867 MeV, respectively. 
We show in Fig.~\ref{fig} the calculated in-medium mass of $J/\psi$  
as a function of baryonic density ratio, $\rho_B/\rho_0$ ($\rho_0 = 0.15$ fm$^{-3}$). 
In cold nuclear matter ($T=0$ and $f_s$ = 0)  
the mass shift $\Delta m_{J/\psi}$ is -3.5 MeV at $\rho_B/\rho_0 = 1.0$ 
{ (normal nuclear matter density).}
The density dependence of $J/\psi$ mass  is through the drop of $D$ meson mass 
calculated using chiral SU(3) plus QCD sum-rule approach~\cite{rahul}. 

For a given baryonic density and at finite temperature, the magnitude of { downward} shift
in the $J/\psi$ mass increases as a function of strangeness 
fraction of the medium. 
This can be understood on the basis that the pseudoscalar $D$ meson mass drops 
more in the strange hadronic medium as the strangeness fraction $f_s$ increases  
(along with nucleons) as discussed in Ref.~\cite{rahul}.
The drop in the mass of the $J/\psi$ meson { as increasing $f_s$} indicates 
that the strange hadronic medium provides an attraction to $J/\psi$ state. 
Furthermore, an impact of finite temperature gives opposite effect  
to that of the strangeness fraction. 
Namely, finite temperature causes an increase of the $J/\psi$ mass. 
For example, { $T = 100$ MeV} at normal nuclear matter density ($\rho_B/\rho_0 = 1.0$ and 
$f_s$=0), the value of $\Delta m_{J/\psi}$ is -2.9 MeV, 
{ while at $T = 0$ it is -3.5 MeV.} 
 
This can be understood in terms of the temperature dependence of pseudoscalar 
$D$ meson { as follows.}  
Since the mass of $D$ meson increases as temperature of the medium 
increases~\cite{rahul}, this in turn gives less contribution for the 
$D$-meson loop contribution to the $J/\psi$ self-energy than that in free space, 
results in to increase the mass of the $J/\psi$ state. 
This indicates that hot medium with zero baryon density ($T = 100$ MeV and $\rho_B = 0$)   
provides repulsive potential, whereas the cold medium ($T = 0$) provides 
attractive potential for the $J/\psi$ state.
 
In the literature the $J/\psi$ mass shift was evaluated using various methods. 
For example, in Ref.~\cite{ko} the authors observed a small positive mass shift 
at normal nuclear matter density when only the effect of the $D$-meson loop was taken into 
account, where the in-medium mass of $D$ meson was used to include the medium effect  
using the momentum-dependent $J/\psi$-$DD$ coupling. 
On the other hand, they obtained a negative mass shift by 
the leading order QCD calculation alone.  
As the total sum of them, they got a slight negative mass shift of the $J/\psi$ meson.    
In Ref.~\cite{tsu} authors also observed a relatively larger negative mass { shift}  
of the $J/\psi$ state by considering the in-medium mass of $D$ meson, 
calculated by the quark-meson coupling (QMC) model.
Although these studies were made in cold nuclear matter, 
we have studied the $J/\psi$ mass shift further in hot and strange hadronic medium 
in addition to the existing studies { made} in cold nuclear matter.  
The  observed in-medium mass of the $J/\psi$ state { in this study} should be helpful, 
to understand the $J/\psi$ suppression, observed in NA50 collaboration 
at 158 GeV/nucleon in Pb-Pb collisions.
We have obtained a moderate negative mass shift of the $J/\psi$ state, 
relative to that observed for the pseudoscalar $D$ meson~\cite{rahul}.
{ The negative mass shift of the $D$ meson is more pronounced than 
that of the $J/\Psi$.}

\section{Conclusion}
We have observed a moderate negative mass shift of the $J/\psi$ state,  
using an effective Lagrangian approach focusing on the 
effect of the $D$ meson loop contribution on the $J/\psi$ self-energy 
in medium as well as in free space. 
The medium effect was taken through the in-medium mass modification of $D$ meson. 
The observed negative mass shift of the $J/\psi$ meson that relatively smaller 
than that of the $D$ ($\bar{D}$) meson, suggests that $J/\psi$ suppression 
will be enhanced by the enhanced dissociation into the $D\bar{D}$ 
meson pair in hadronic medium, due to the downward shift of the threshold. 
The results of the present study, together with more detailed study, 
should be helpful for better understanding of future HICs, such as CBM and PANDA 
under FAIR facility. In particular, to study further the effects 
of temperature and strangeness fraction, will add important, 
new information. We plan to do an elaborated study focusing on these 
issues in the future. 

\section{Acknowledgements}
R.C. thanks Arvind Kumar (National Institute of Technology Jalandhar) 
for useful discussions. 
{ 
K.T. was supported by the Conselho Nacional de Desenvolvimento
Cient\'{i}fico e Tecnol\'{o}gico (CNPq)
Process, No.~313063/2018-4, and No.~426150/2018-0,
and Funda\c{c}\~{a}o de Amparo \`{a} Pesquisa do Estado
de S\~{a}o Paulo (FAPESP) Process, No.~2019/00763-0,
and was also part of the projects, Instituto Nacional de Ci\^{e}ncia e
Tecnologia -- Nuclear Physics and Applications (INCT-FNA), Brazil,
Process. No.~464898/2014-5, and FAPESP Tem\'{a}tico, Brazil, Process,
No.~2017/05660-0.
}

\end{document}